\documentclass[aps,pra,twocolumn,superscriptaddress,nofootinbib]{revtex4}
\usepackage{amsthm}
\usepackage{amsmath}
\usepackage{amssymb}
\usepackage{graphicx}
\usepackage{url}
\usepackage{dsfont}
\usepackage{float}
\usepackage{bm}
\usepackage{ifthen}
\usepackage[usenames,dvipsnames]{color}
\usepackage{mathrsfs}
\usepackage[colorlinks=true,citecolor=blue,urlcolor=black]{hyperref}
\usepackage{float}
\usepackage{subfigure}
\usepackage{enumitem}
\usepackage{color}
\usepackage{times}

\newcommand{\eg}{{{e.g.}}}
\newcommand{\ie}{{{i.e.}}}

\newcommand{\tn}[1]{\textnormal{#1}}
\newcommand{\be}{\begin{equation}}
\newcommand{\ee}{\end{equation}}

\newcommand{\leak}{\tn{leak}_{\tn{EC}}}
\newcommand{\epscorr}{\eps_{\tn{cor}}}
\newcommand{\esec}{\eps_{\tn{sec}}}

\newcommand{\sket}[1]{{\ensuremath{\lvert#1\rangle}}}
\newcommand{\lket}[1]{{\ensuremath{\left\lvert#1\right\rangle}}}
\newcommand{\ket}[1]{\if@display\lket{#1}\else\sket{#1}\fi}
\newcommand{\sbra}[1]{{\ensuremath{\langle#1\rvert}}}
\newcommand{\lbra}[1]{{\ensuremath{\left\langle#1\right\rvert}}}
\newcommand{\bra}[1]{\if@display\lbra{#1}\else\sbra{#1}\fi}
\newcommand{\sbraket}[2]{{\ensuremath{\langle#1\rvert#2\rangle}}}
\newcommand{\lbraket}[2]{{\ensuremath{\left\langle#1\!\left\rvert\vphantom{#1}#2\right.\!\right\rangle}}}
\newcommand{\braket}[2]{\if@display\lbraket{#1}{#2}\else\sbraket{#1}{#2}\fi}

\newcommand{\sketbra}[2]{{\ensuremath{\lvert #1\rangle\!\langle #2\rvert}}}
\newcommand{\lketbra}[2]{{\ensuremath{\left\lvert #1\right\rangle\!\!\left\langle #2\right\rvert}}}
\newcommand{\ketbra}[2]{\if@display\lketbra{#1}{#2}\else\sketbra{#1}{#2}\fi}

\newcommand{\eps}{\varepsilon}

\theoremstyle{plain}

\theoremstyle{definition}

\begin{document}
\title{Experimental quantum key distribution with source flaws}

\author{Feihu Xu}
 \email{feihu.xu@utoronto.ca}
\thanks{Present address: Research Laboratory of Electronics, Massachusetts Institute of Technology, 77 Massachusetts Avenue, Cambridge, Massachusetts 02139, USA}
\affiliation{Centre for Quantum Information and Quantum Control (CQIQC), Dept. of Electrical \& Computer Engineering and Dept. of Physics, University of Toronto, Toronto,  Ontario, M5S 3G4, Canada}

\author{Kejin Wei}
\affiliation{Centre for Quantum Information and Quantum Control (CQIQC), Dept. of Electrical \& Computer Engineering and Dept. of Physics, University of Toronto, Toronto,  Ontario, M5S 3G4, Canada}
\affiliation{School of Science and State Key Laboratory of Information Photonics and Optical Communications, Beijing University of Posts and Telecommunications, Beijing 100876, China}

\author{Shihan~Sajeed}
\affiliation{Institute for Quantum Computing (IQC), University of Waterloo, Waterloo, ON, N2L~3G1 Canada}
\affiliation{Dept. of Electrical and Computer Engineering, University of Waterloo, Waterloo, ON, N2L~3G1 Canada}

\author{Sarah~Kaiser}
\affiliation{Institute for Quantum Computing (IQC), University of Waterloo, Waterloo, ON, N2L~3G1 Canada}
\affiliation{Dept. of Physics and Astronomy, University of Waterloo, Waterloo, ON, N2L~3G1 Canada}

\author{Shihai Sun}
\affiliation{College of Science, National University of Defense Technology, Changsha 410073, P.R.China}

\author{Zhiyuan Tang}
\affiliation{Centre for Quantum Information and Quantum Control (CQIQC), Dept. of Electrical \& Computer Engineering and Dept. of Physics, University of Toronto, Toronto,  Ontario, M5S 3G4, Canada}

\author{Li Qian}
\affiliation{Centre for Quantum Information and Quantum Control (CQIQC), Dept. of Electrical \& Computer Engineering and Dept. of Physics, University of Toronto, Toronto,  Ontario, M5S 3G4, Canada}

\author{Vadim~Makarov}
\affiliation{Institute for Quantum Computing (IQC), University of Waterloo, Waterloo, ON, N2L~3G1 Canada}
\affiliation{Dept. of Electrical and Computer Engineering, University of Waterloo, Waterloo, ON, N2L~3G1 Canada}
\affiliation{Dept. of Physics and Astronomy, University of Waterloo, Waterloo, ON, N2L~3G1 Canada}

\author{Hoi-Kwong Lo}
\affiliation{Centre for Quantum Information and Quantum Control (CQIQC), Dept. of Electrical \& Computer Engineering and Dept. of Physics, University of Toronto, Toronto,  Ontario, M5S 3G4, Canada}

\date{\today}
\begin{abstract}
Decoy-state quantum key distribution (QKD) is a standard technique in current quantum cryptographic implementations. Unfortunately, existing experiments have two important drawbacks: the state preparation is assumed to be perfect without errors and the employed security proofs do not fully consider the finite-key effects for general attacks. These two drawbacks mean that existing experiments are not guaranteed to be secure in practice. Here, we perform an experiment that for the first time shows secure QKD with imperfect state preparations over long distances and achieves rigorous finite-key security bounds for decoy-state QKD against coherent attacks in the universally composable framework. We quantify the source flaws experimentally and demonstrate a QKD implementation that is tolerant to channel loss despite the source flaws. Our implementation considers more real-world problems than most previous experiments and our theory can be applied to general QKD systems. These features constitute a step towards secure QKD with imperfect devices.
\end{abstract}
\maketitle

Quantum key distribution (QKD), offering information-theoretic security in communication, has aroused great interest among both scientists and engineers~\cite{QKDreview1,QKDreview2,QKDreview3}. Commercial systems have already appeared on the market and various QKD networks have been developed. The most important question in QKD is its security. This fact has finally been proven in a number of important papers~\cite{Lo:1999,Shor:2000,GLLP:2004} (see~\cite{QKDreview2} for a review on this topic). However, for real-life implementations that are mainly based on attenuated laser pulses, the occasional production of multi-photons and channel loss make QKD vulnerable to various subtle attacks, such as the photon-number-splitting attack~\cite{PNS:2000}. Fortunately, the decoy-state method~\cite{Hwang:2003, Lo:2005, Wang:2005} has solved this security issue and dramatically improved the performance of QKD with faint lasers. Several experimental groups have demonstrated that decoy-state BB84 is secure and feasible under real-world conditions~\cite{zhao2006experimental,rosenberg2007long,peng2007experimental,schmitt2007experimental,dixon2008gigahertz}. As a result, decoy-state method has become a standard technique in many current QKD implementations~\cite{lucamarini2013efficient,nauerth2013air,Lo:MDIQKD,MDIQKD:review,liu2013experimental,rubenok2013real,da2013proof,Tang2014polarization,Tang200km}.

Until now, QKD experiments~\cite{zhao2006experimental,rosenberg2007long,peng2007experimental,schmitt2007experimental,dixon2008gigahertz,lucamarini2013efficient,nauerth2013air,Lo:MDIQKD,MDIQKD:review,liu2013experimental,rubenok2013real,da2013proof,Tang2014polarization,Tang200km}  have had two important drawbacks. The first one is that in the key rate formula of all existing experiments, it is commonly assumed that the phase/polarization encoding is done \emph{perfectly} without errors. Thus, the state preparation is assumed to be basis-independent, i.e. the density matrices for the two conjugate basis are assumed to be the same. These are highly unrealistic assumptions and may mean that the key generation is actually not proven to be secure in a real QKD experiment. What if we use a key rate formula that takes imperfect encodings into account? Standard Gottesman-Lo-L\"{u}tkenhaus-Preskill (GLLP) security proof~\cite{GLLP:2004} (see also~\cite{Mar2010, Woodhead2013}) does allow one to do so. Unfortunately, the key rate will be reduced substantially because the GLLP formalism is very conservative and the resulting protocol is not tolerant to channel loss~\cite{tamaki2012phase}. We remark that source flaw is a serious concern in not only decoy-state BB84 but also measurement-device-independent QKD~\cite{Lo:MDIQKD,MDIQKD:review}, quantum coin flipping~\cite{berlin2011experimental,pappa2014experimental} and blind quantum computing~\cite{dunjko2012blind}.

To address the source flaw problem, Tamaki et al. put forward a proposal~\cite{tamaki2013loss}, which allows QKD protocols that are tolerant to channel loss despite the source flaws. We call it a \emph{loss-tolerant protocol}. The key insight is that as long as the single-photon components of the four BB84 states remain inside a two-dimensional Hilbert space (which we call a \emph{qubit assumption}), Eve can not enhance state-preparation flaws by exploiting the channel loss and Eve's information can be bounded by the rejected data analysis~\cite{rejectedanalysis}. Nevertheless, Ref.~\cite{tamaki2013loss} is only valid in the asymptotic limit with an infinite number of signals and decoy states, and thus it has a number of important limitations when it is applied in practice. These limitations include: (i) How to extend it to the practical case with only a finite number of types of decoy states? (ii) How to extend it to the case with a finite number of transmitted signals (which is normally called finite-key analysis)? (iii) How to verify the qubit assumption made in the theory? (iv) How to quantify the source flaws in practice? (v) How to implement the loss-tolerant protocol in experiment? In this paper, we overcome these five limitations (see discussions below).

The second drawback in previous experiments is that the finite-key security claims were made with the assumption that the eavesdropper (Eve) was restricted to particular types of attacks (\eg, collective attacks) or that the finite-key analysis was not rigorous (\eg, the security did not satisfy the universally composable security definition~\cite{ben2005universal,renner2005universally}). Unfortunately, such assumptions cannot be guaranteed in practice. While Ref.~\cite{bacco2013experimental} has reported an attempt implementing the rigorous finite-key analysis proposed in~\cite{tomamichel2012tight}, a slight drawback is that both the theory and experiment assume a perfect single-photon source without decoy states. Very recently, Lim et al. provide, for the first time, tight and rigorous security bounds against general quantum attacks (i.e., coherent attacks) for decoy-state QKD~\cite{lim2013concise} (see also~\cite{hayashi2013security}). These bounds are obtained by combining the finite-key analysis of~\cite{tomamichel2012tight} and the finite-data analysis of~\cite{curty2013finite}. Nonetheless, Ref.~\cite{lim2013concise} still assumes basis-independent state preparation. A theory that removes the assumption of basis independency, and a QKD experiment that implements such an advanced theory have not been reported yet.

In this paper, we offer a link between the theory and experiment to consider both source flaws and finite-key effect in practical QKD. We overcome the limitations in the loss-tolerant protocol and implement this protocol in experiment. The advances of our work are both theoretical and experimental. On the theoretical side, our contributions are as follows. First, we provide both a finite-key analysis and a practical decoy state method for the loss-tolerant protocol, thus making this protocol applicable in a real experiment. Our parameter estimation method considers general source flaws and does not rely on the assumption of basis independency of prepared states. Second, we perform a detailed simulation for the loss tolerant protocol and show that this protocol can substantially outperform GLLP in a practical setting with a reasonable data-set. We note in passing that the loss-tolerant protocol only requires three states for the security analysis, thus it can simplify conventional BB84 implementations, especially for those based on four laser sources~\cite{peng2007experimental, schmitt2007experimental,nauerth2013air}, where one could keep one laser just as back-up in case certain laser fails, without any decrease in performance. Third, we perform a comprehensive analysis on the qubit assumption in a standard one-way phase-encoding system and have verified such assumption with high accuracy by using standard optical devices.

On the experimental side, by modifying a commercial plug\&play QKD system, we perform the first QKD demonstration considering source flaws. We quantify these flaws experimentally and include them in the key rate formula. Based on the loss-tolerant protocol, we successfully generate secure keys over different channel lengths, up to 50 km standard telecom fibers. In contrast, not even a single bit of secure key can be extracted with GLLP security proof. Moreover, in our implementation, we apply a tight finite-key analysis -- that does not rely on the assumption of basis-dependent state preparation -- to generate keys, which are secure against the most powerful (coherent) attacks in the universally composable framework. We emphasize that our implementation, security analysis and parameter estimation procedure can be applied to general discrete-variable QKD systems. Very recently, similar progress on the continuous-variable QKD system has been reported in~\cite{GehringCVQKD}.


\subsection*{Theory}
\emph{Three-state QKD:} The loss-tolerant protocol is a general method that works not only for the standard BB84 protocol, but even for the three-state protocol~\cite{boileau2005unconditional,fung2006security} where there is a strong asymmetry between the two bases. The loss-tolerant protocol includes basis mismatch events for security analysis to beat eavesdroppers. The three-state QKD runs almost the same as BB84 except that: i) Alice sends Bob only three pure states \{$\ket{0_{z}}$, $\ket{0_{x}}$, $\ket{1_{z}}$\}, where $\ket{i_{j}}$ ($i\in$\{0,1\} and $j\in$\{$\mathsf{Z}$, $\mathsf{X}$\}) denotes the state associated with bit ``$i$" in $j$ basis; ii) the rejected data (\ie, the detection events when Alice and Bob use different basis) are used for the estimation of the phase error rate~\cite{rejectedanalysis}. Based on the security analysis with biased basis choice, Alice and Bob can generate a secret key only from those instances where both of them select the $\mathsf{Z}$ basis~\cite{tamaki2013loss}.

\emph{The qubit assumption and its verification:} The qubit assumption is normally required in the security proofs~\cite{QKDreview2} to avoid subtle attacks such as unambiguous state discrimination attack~\cite{USDattack}. With the qubit assumption in place, using large deviation techniques (e.g. Hoeffding's inequality~\cite{Hoeffding} or quantum de Finetti theorem~\cite{RennerThesis}), one can show that effectively Eve can only apply the same super-operator on each transmitted qubit. This greatly simplifies the security proofs. In practice, however, no previous works have verified this assumption. Note that a specific attack to exploit the higher dimensionality of state preparation has been proposed in~\cite{sun2011passive} recently. Here we have verified that the qubit assumption can be made valid (to a large degree) in practice, while further work needs to be done to make it more rigorous. The detailed results are shown in Supplementary Material.

\emph{Finite-key analysis:} So far, the loss-tolerant protocol was only proven in the asymptotic case, \ie, the legitimate users have unlimited resources~\cite{tamaki2013loss}. Such an asymptotic case is impossible in practice. Here, to implement the loss-tolerant protocol, we extend it to a general practical setting with finite keys and finite decoy states by synthesising~\cite{tamaki2013loss} and~\cite{tomamichel2012tight,lim2013concise}. The $\esec$-secret key length in the $\mathsf{Z}$ basis is given by
\begin{eqnarray} \label{eqn1}
\ell \geq s_{z,0}^{L}+s_{z,1}^{L}[q - {h}\left( e_{x,1}^{U} \right)] -\leak \\ \nonumber
- 6\log_2\frac{21}{\esec}-\log_2\frac{2}{\epscorr},
\end{eqnarray}
where~$h(y)$=$-y\log_2y-(1-y)\log_2(1-y)$~is the binary entropy function; $s_{z,0}^{L}$, $s_{z,1}^{L}$ and $e_{x,1}^{U}$ are the lower bound of vacuum events, the lower bound of single-photon events, and the upper bound of the phase error rate, associated with the single-photon events in $\mathsf{Z}$ basis, respectively; $q$ is the maximum fidelity between states prepared in $\mathsf{Z}$ basis and states prepared in $\mathsf{X}$ basis, which characterizes the quality of the source~\cite{tomamichel2012tight}; $\leak=n_{z,\mu}f_{e}h\left(e_{z} \right)$ is the size of the information exchanged during error-correction, where $n_{z,\mu}$ and $e_z$ denote respectively the gain counts for signal state and quantum bit error rate (QBER) and $f_{e}\geq 1$ is the error correction inefficiency function (we choose $f_{e}=1.16$ in this paper); $6\log_2\frac{21}{\esec}$ and $\log_2\frac{2}{\epscorr}$ are respectively the secrecy and correctness parameter; $\ell$ quantifies the lower bound of final key length and the key rate (per optical pulse) is given by $R^{L}$=$\ell/N$ with $N$ denoting the total number of signals (optical pulses) sent by Alice. This key formula uses a security proof that is based on an uncertainty relation for smooth entropies~\cite{tomamichel2012tight} and it fulfills the composable security definition~\cite{ben2005universal,renner2005universally}.

\emph{Finite decoy-state protocol:} In practice, $s_{z,0}^{L}$, $s_{z,1}^{L}$ and $e_{x,1}^{U}$ are estimated using the decoy-state method. Here, we propose a novel method for the estimation of the phase error rate $e_{x,1}^{U}$. In our analysis, besides the signal state $\mu$, we consider two additional decoy states, $\nu$ and $\omega$, where $\mu$, $\nu$ and $\omega$ are the mean photon numbers of weak coherent pulses and they satisfy $\mu > \nu > \omega \geq 0$.  Hence, the intensity setting $k\in \{\mu,\nu,\omega\}$. The key novelty to estimate $e_{x,1}^{U}$ is obtained by estimating the transmission rate of a virtual quantum signal sent by Alice (see Supplementary Material). The estimation of this transmission rate uses the rejected detection counts~\cite{rejectedanalysis}, \ie, considering the detection events associated with single photons when Alice and Bob use different bases. By doing so, we have the key advantage of removing the assumption of basis-independent state preparation entirely. The estimation result is shown in Eq.~(\ref{A:eqn4}) of Methods. $s_{z,0}^{L}$ and $s_{z,1}^{L}$ can be estimated using a method similar to~\cite{lim2013concise}, from the detection events $n_{z,k}$. See Methods for the details of our decoy state protocol.

\subsection*{Experiment}
\begin{figure*}[!t]
\centering
\resizebox{14cm}{!}{\includegraphics{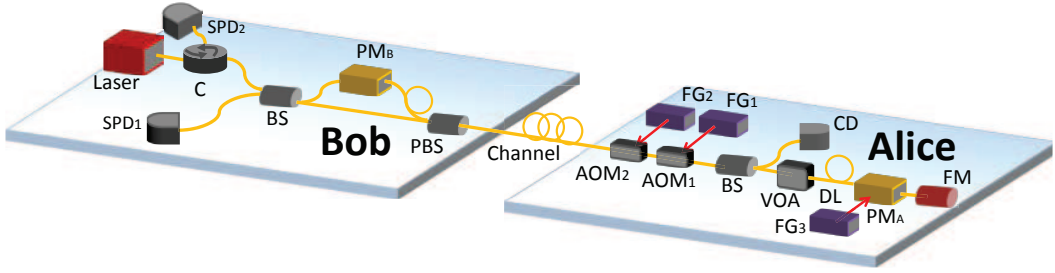}} \caption{Experimental setup. SPD$_{1}$/SPD$_{2}$, single-photon detector; C, circulator; PM$_{A}$/PM$_{B}$, phase modulator; BS, beam splitter; PBS, polarization beam splitter; CD, classical photo-detector; VOA, variable optical attenuator; AOM$_{1}$/AOM$_{2}$, acousto-optic modulator; FG, function generator; DL, delay line; FM, Faraday mirror. PM$_{A}$, controlled by FG$_{3}$, randomly selects a phase from \{0, $\pi/2$, $\pi$\} for the experiment. AOM$_{1}$ randomly modulates the intensity of each pulse to be either signal state level or decoy state level, while AOM$_{2}$ compensates the phase shift due to AOM$_{1}$.} \label{Fig:setup}
\end{figure*}

\emph{System description:} We implement the loss-tolerant protocol with a modified commercial ID-500 plug\&play QKD system (manufactured by ID~Quantique)~\cite{stucki2002quantum}. Nonetheless, we remark that our methods of parameter optimizations, finite key analysis, the quantification of phase modulation errors and the implementation of loss-tolerant protocol can also be applied to one-way QKD systems. Here, we use the plug\&play QKD system simply as an example to illustrate our \emph{general} methods.

The initial plug\&play system employs the phase-coding QKD scheme and it works as follows (see Fig.~\ref{Fig:setup})~\cite{IDQ}. Bob first sends two laser pulses (\ie, signal and reference pulse) to Alice. Alice uses the reference pulse as a synchronization signal (detected by her classical photo-detector) to activate her phase modulator (PM). Then Alice modulates the phase of the signal pulse only, attenuates the two pulses to single photon level, and sends them back to Bob. Bob randomly chooses his measurement basis by modulating the phase of the returning reference pulse and detects the interference signals with his two single-photon detectors (SPDs).

Now, we present our modifications on top of ID-500 in order to realize the loss-tolerant protocol with decoy states. To implement the decoy-state protocol, we add two acousto-optic modulators (AOMs, Brimrose) to achieve polarization-insensitive intensity modulation. AOM$_{1}$ -- driven by a waveform with random patten generated from a function generator (FG$_{1}$, Agilent 88250A) -- is used for the decoy modulation, while AOM$_{2}$ -- driven by a fixed waveform generated from FG$_{2}$ -- is used to compensate the phase shift caused by the frequency shift of the AOM~\cite{zhao2006experimental}. To implement the three-state protocol, we adopt another FG, i.e., FG$_{3}$, to control the phase modulation of PM$_{A}$. FG$_{1}$ and FG$_{3}$ are loaded with random numbers generated from a quantum random number generator~\cite{XuQRNG}. We have measured the main system parameters as shown in Table~\ref{Tab:idq}.

\begin{table}\center
\begin{tabular}{c @{\hspace{0.3cm}} c @{\hspace{0.3cm}} c  @{\hspace{0.3cm}} c @{\hspace{0.3cm}} c @{\hspace{0.3cm}} c}
\hline \hline $\lambda$ & $e_{d}$ & $\eta_{Bob}$ & $Y_0$ & \emph{f} \\
\hline $1551.71$ nm & $2.35\%$ & $5.05\%$ & $4.01\times10^{-5}$ & 5 MHz \\
\hline \hline
\end{tabular}
\caption{Parameters measured in ID-500 commercial QKD system, including laser wavelength $\lambda$, optical misalignment error $e_{d}$ (the probability that a photon hits the erroneous detector), Bob's overall quantum efficiency $\eta_{Bob}$, dark count rate per pulse $Y_{0}$ for each detector and system repetition rate $f$. }\label{Tab:idq}
\end{table}

\begin{table}[ht!]
\centering
\begin{tabular}{c  @{\hspace{0.3cm}} c @{\hspace{0.3cm}} c @{\hspace{0.3cm}} c @{\hspace{0.3cm}} c }
  \hline \hline
   System & $\theta$ & $D_{1,\theta}$ & $D_{2,\theta}$  & $\bar{\delta}_{\theta}$\\
   \hline
  ID-500 & 0 & 630 & 867678 & -  \\
         & $\pi/2$ & 456735 & 444336 & 0.013  \\
         & $\pi$ & 856245 & 4744 & 0.134  \\
         & $3\pi/2$ & 464160 & 436962 & 0.030  \\
  \hline
  Clavis2 & 0 & 727 & 1075320 & -  \\
         & $\pi/2$ & 546724 & 527735 & 0.023  \\
         & $\pi$ & 1111574 & 6990 & 0.145  \\
         & $3\pi/2$ & 566813 & 531417 & 0.037  \\
  \hline \hline
\end{tabular}
\caption{Raw counts and modulation errors for Alice's phase modulator in ID-500 and Clavis2 commercial plug\&play systems. $D_{1,\theta}$ ($D_{2,\theta}$) represents the detections counts of SPD$_{1}$ (SPD$_{2}$). $\bar{\delta}_{\theta}$, given by Eq.~(\ref{Eqn:deltaupp}), is the upper bound of modulation error for a given phase $\theta$.} \label{Tab:exp:calibration}
\end{table}

\emph{Quantifying modulation error:} We quantify the modulation error $\delta_{\theta}$ in the source through calibrating Alice's PM, a LiNbO$_{3}$ waveguide based electro-optical modulator, on two plug\&play QKD systems -- ID 500 and Clavis2~\cite{IDQ}. $\delta_{\theta}$ is defined as the difference between the actual phase and the expected phase $\theta\in$\{0, $\pi/2$, $\pi$\, $3\pi/2$\}. We find that in ID-500, the voltages \{0, 0.30$V_{m}$, 0.62$V_{m}$, 0.92$V_{m}$\} modulate the expected phases \{0, $\pi/2$, $\pi$\, $3\pi/2$\}, where $V_{m}\approx$ 3.67 V is a maximal value allowed on Alice's PM. The calibration process is as follows. Alice is directly connected to Bob with a short fiber (about 1 m), Alice scans the voltages applied to her PM, Bob sets his own PM at a fixed unmodulated phase \{0\} and records the detection counts of his two SPDs. These counts are denoted by $D_{1,\theta}$ and $D_{2,\theta}$. The detections counts on ID-500 and Clavis2 are shown in Table~\ref{Tab:exp:calibration}.

In ID-500, to quantify $\delta_{\theta}$, we first determine the detector efficiencies ($\eta_{d1}$, $\eta_{d2}$) and the dark count rates ($Y_{0,d1}$, $Y_{0,d2}$) for Bob's two SPDs and find that $\eta_{d1}=5.05\%$ and $\eta_{d2}=4.99\%$ and $Y_{0,d1}\approx Y_{0,d2}=4.01\times10^{-5}$. In Table~\ref{Tab:exp:calibration}, $D_{1,0}$ quantifies the amount of global misalignment between Alice and Bob (i.e. the summation of the dark counts and the imperfect visibility). This global misalignment can increase QBER, but it is irrelevant to bound Eve's information in the loss-tolerant protocol~\cite{tamaki2013loss}. Only the relative orientation between the three states prepared by Alice quantifies the source flaws that can be potentially exploited by Eve. Hence, we subtract $D_{1,0}$ in the quantification of $\delta_{\theta}$. In our analysis of the statistics, we use Hoeffding's inequality~\cite{Hoeffding} to guarantee the definition of composable security. The upper bound of $\delta_{\theta}$ is given by:
\begin{widetext}
\begin{equation} \label{Eqn:deltaupp}
\begin{aligned}
\delta_{\theta} \leq \bar{\delta}_{\theta} = |\theta-2\arctan(\sqrt{\frac{((D_{1,\theta}+\Delta(D_{1,\theta},\eps))-(D_{1,0}-\Delta(D_{1,0},\eps)))/\eta_{d1}}
{((D_{2,\theta}-\Delta(D_{2,\theta},\eps))-(D_{1,0}+\Delta(D_{1,0},\eps)))/\eta_{d2}}})|,
\end{aligned}
\end{equation}
\end{widetext}
where $\Delta(D_{i,\theta},\eps)$=$\sqrt{D_{i,\theta}/2\log(1/\eps)}$ (with $i\in\{0,1\}$)~\cite{Hoeffding}. In general, if $Y_{0,d1}\neq Y_{0,d2}$ in a practical system, in Eq.~(\ref{Eqn:deltaupp}), we can use $D_{i,\theta}$ to subtract the dark counts of detector $d_{i}$. Here, we choose a failure probability $\eps=10^{-10}$ (i.e. a confidence level $1-2\times10^{-10}$). The upper bounds of $\delta_{\theta}$ are shown in Table~\ref{Tab:exp:calibration}. From this table, the error $\delta$ in ID-500 is upper bounded by the case of $\delta_{\pi}$, \ie, $\delta \leq \bar{\delta}_{\pi}=0.134$.

Using the same method for Clavis2, we find that $\delta$ is upper bounded by $\delta \leq \bar{\delta}_{\pi}=0.145$. Notice that $\delta$ can also be estimated using the interference visibility or the extinction ratio of the PM~\cite{tamaki2012phase}. In a system with an advanced phase-stabilized interferometer~\cite{honjo2004differential}, the value of $\delta\leq0.062$ corresponds to about 99.9\% visibility or 30 dB extinction ratio.

\emph{Implementation of loss-tolerant protocol:} In our demonstration, we implement the loss-tolerant protocol over standard fibre lengths (L) of 5, 20 and 50 km. In the 5 and 20 km experiments, we performed a real decoy-state QKD implementation with optimized parameters. We use FG1 to randomly modulate the signal and decoy states and use FG3 to randomly modulate the three states of \{$\ket{0_{z}}$, $\ket{0_{x}}$, $\ket{1_{z}}$\}. Before the experiment, we performed a numerical simulation to optimize the implementation parameters. Our optimization routine is similar to~\cite{ma2005practical}, while the difference is that we use the rigorous finite-key security bounds (see Eq.~(\ref{eqn1})) to predict the key rate. The optimal parameters are shown in Table~\ref{Tab:exp:results}, which include intensities of $\mu$ (signal), $\nu$ (decoy), $\omega$ (vacuum), intensity-probabilities of $P_{\mu}$, $P_{\nu}$, $P_{\omega}$ ($P_{\omega}=1-P_{\mu}-P_{\nu}$), and basis-probabilities of $P_{z}$ and $P_{x}$ (which are identical for Alice and Bob). In the 50 km experiment, we removed the two AOMs due to their high loss (over 3 dB each) and used the VOA in Alice to modulate the decoy intensities for a proof of concept decoy-state modulation.

\begin{table*}[ht!]
\centering
\begin{tabular}{c c | c  c  c  c  c  c |  c  c  c | c  c  c}
  \hline \hline
    Channel & & & &Parameters& & & && Estimation& & & Performance&  \\
   \hline
   L (km) & Attn (dB) & N & $\mu$ & $\nu$  & $P_{\mu}$ & $P_{\nu}$ & $P_{z}$ & $s_{z,0}^{L}$ & $s_{z,1}^{L}$ & $e_{x,1}^{U}$ & $e_z$ & $l$ & $R^{L}$ \\
   \hline
   5 & 1.4 & $7.84\times10^{9}$ &  0.41 & 0.05 & 0.64 & 0.27 & 0.70 & $7.40\times10^{4}$ & $3.02\times10^{7}$ & 6.28\%  & 2.67\% & $1.06\times10^{7}$ & $1.40\times10^{-3}$ \\

   20 & 4.5 & $7.84\times10^{9}$ & 0.37 & 0.06  & 0.40 & 0.50 & 0.60 & $6.15\times10^{4}$ & $6.58\times10^{6}$ & 8.67\% & 2.74\% & $8.07\times10^{5}$ & $1.03\times10^{-4}$ \\

   50 & 10.5 & $5.23\times10^{10}$ & 0.55 & 0.06  & 0.74 & 0.18 & 0.50 & $3.36\times10^{5}$ & $1.33\times10^{7}$ & 8.46\% & 2.98\% & $1.07\times10^{6}$ & $2.14\times10^{-5}$ \\
  \hline \hline
\end{tabular}
\caption{\textbf{Implementation parameters and experimental results.} $N$ is the total number of pulses sent by Alice. $P_{\mu}$, $P_{\nu}$, $P_{\omega}=1-P_{\mu}-P_{\nu}$ are the probabilities to choose different intensities. $P_{z}$ and $P_{x}=1-P_{z}$ are the probabilities to choose the two bases. $\omega$ is about 0.001 for 5 and 50 km experiments, and it is about 0.003 for 20 km experiment. The estimation results are obtained by plugging the experimental counts, shown in Supplementary Material, into the decoy-state estimation equations shown in Methods. The key rate is obtained from Eq.~(\ref{eqn1}).} \label{Tab:exp:results}
\end{table*}

\emph{Experimental results:} Our measurement and post-processing are different from previous experiments in that we directly measure the detection \emph{counts} instead of detection probabilities (so-called gains in former experiments~\cite{zhao2006experimental,rosenberg2007long, peng2007experimental, schmitt2007experimental, dixon2008gigahertz, lucamarini2013efficient}) and we also record the basis-mismatch counts. In the 5 km and 20 km experiments, we chose to operate the system for a few hours and collected about 75 sets of data, with each set of about 104.5 million pulses, which corresponds to a total number of pulses $N=7.84\times10^{9}$. In the 50 km experiment, we collected about 500 sets of data and sent a total number of $N=5.23\times10^{10}$ pulses. The details of these experimental counts are shown in Supplementary Material.

\begin{figure}[!t]
\centering
\resizebox{8cm}{!}{\includegraphics{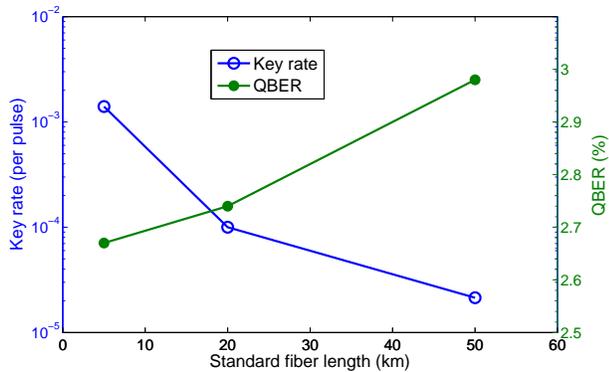}} \caption{Experimental final secret key rate (blue circle) and QBER (green dot) versus distance.} \label{Fig:expresults}
\end{figure}

In our analysis of experimental data, we consider a conservative security parameter (i.e. the summation of all failure probabilities) $\epsilon_{tot}=10^{-10}$. From the model of Eq.~(\ref{Eqn:state}) and the modulation errors of Table~\ref{Tab:exp:calibration}, we find that $q=0.79$ (see Eq.~(\ref{eqn1})). By plugging the experimental counts into the decoy-state estimations (see Methods) and using Eq.~(\ref{eqn1}), we obtain the experimental results listed in Table~\ref{Tab:exp:results} and Fig.~\ref{Fig:expresults}. The system's QBER is below 3\%. Based on the loss-tolerant analysis, a secure key rate (per optical pulse) of $1.40\times10^{-3}$ was generated at 5 km, while at 50 km it was $2.14\times10^{-5}$. Given the 5 MHz repetition rate, the key rates per second are 7 kbps and 107 bps respectively. Over 1 kilobit of unconditionally secure keys are exchanged between Alice and Bob. The security of these keys considers source flaws and satisfies the composable security definition, and it can withstand general attacks by Eve. With state-of-the-art high speed QKD system working at GHz repetition rate~\cite{lucamarini2013efficient}, our loss-tolerant analysis can easily enable a key rate of megabit per second.

As a comparison to previous security analysis (\eg, GLLP), with the source flaws $\delta$=0.134, no matter how many decoy states we choose or how large the data size we use, the key generation rate will hit zero at only about 10 km based on GLLP~\cite{GLLP:2004,tamaki2012phase}. In other words, at 20 km and 50 km, not even a single bit could be shared between Alice and Bob with guaranteed security with previous GLLP security proof. This means that if considering source flaws in previous long-distance decoy-state experiments~\cite{zhao2006experimental,rosenberg2007long,peng2007experimental,schmitt2007experimental,dixon2008gigahertz,lucamarini2013efficient,nauerth2013air}, the key generation might \emph{not} be proven to be secure. In contrast, our analysis with the loss-tolerant protocol can easily achieve high secure key generation rate over long distances even in the presence of source flaws.

\subsection*{Discussion and Conclusion}
\begin{figure}[!t]
\centering
\resizebox{8cm}{!}{\includegraphics{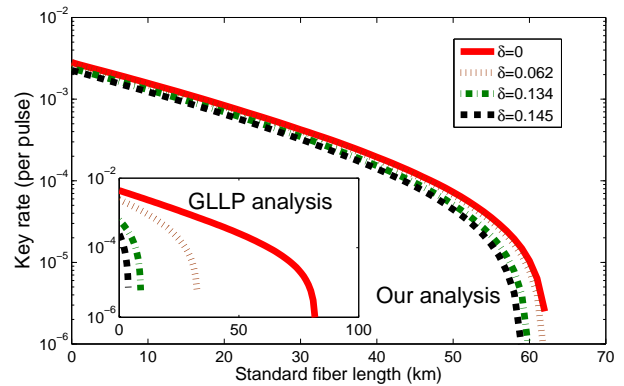}} \caption{Numerical simulation of decoy-state QKD with source flaws in a practical setting. The simulation is conducted with parameters in Table~\ref{Tab:idq}, $N$=$5\times10^{10}$ and $\epsilon_{tot}$=$10^{-10}$. The main figure is for the three-state protocol based on our loss-tolerant security analysis, while the inset figure is for the decoy-state BB84 protocol based on the GLLP security analysis (see Methods for the model). The power of our security analysis is explicitly shown by the fact that GLLP delivers a key rate that decreases rapidly when $\delta$ increases. The maximal tolerant distance is about 9 km for our QKD system (green dashed-dotted curve in the inserted figure). In contrast, our analysis can substantially outperform GLLP and it is loss-tolerant to source flaws. Our QKD set up can be made secure over 60 km and the secure key rate is almost the same as the case without considering source flaws (\ie, assuming $\delta$=0). } \label{Fig:simulation:flaws}
\end{figure}

\emph{Numerical simulation: } With $\delta$ and the parameters in Table~\ref{Tab:idq}, we perform a simulation to numerically study our security analysis in a practical setting. Fig.~\ref{Fig:simulation:flaws} shows the simulation results, where similar to our 50 km experiment, we use $N$=$5\times10^{10}$ and $\epsilon_{tot}$=$10^{-10}$. For comparison, this figure also includes the key rate for the decoy-state BB84 based on the GLLP security analysis (See Methods for the model). The power of our security analysis is explicitly shown by the fact that GLLP delivers a key rate that decreases rapidly when $\delta$ increases. The maximal tolerant distance is about 9 km for our QKD system. This is because GLLP considers the worst case scenario where losses can increase the fidelity flaw~\cite{GLLP:2004,tamaki2012phase}. Our security analysis, however, can substantially outperform GLLP and it is loss-tolerant to source flaws. Our QKD set up can be made secure over 60 km and the secure key rate is almost the same as the case without source flaws.

\emph{Conclusion:}
We have demonstrated decoy-state QKD with imperfect state preparation and employed tight finite-key security bounds with composable security against coherent attacks. By overcoming the limitations of the loss-tolerant protocol and quantifying the source flaws, we take the real-world imperfections that were not addressed before, into the consideration of security analysis to provide enhanced for practical QKD systems. Our work constitutes an important step towards secure QKD with imperfect devices in practice. In our paper, we ignore certain imperfections in the source such as the intensity fluctuations of signal/decoy states, which have a small effect and can be taken care of using previous result~\cite{wang2008general}. Also, we assume that there is no unwanted information leakage from the source. How to protect the source against Eve's active influence will be a subject for future investigations~\cite{shihan}. Moreover, to guarantee the qubit assumption, a device-independent dimension witness could be used to verify the dimension of the source~\cite{gallego2010device}. Furthermore, it will be interesting to work out a refined security proof that include all possible (small) imperfections and side channels in the source and extend our results to MDI-QKD~\cite{Lo:MDIQKD}. Thus, one can solve the problem of not only imperfect source but also remove all loopholes in the detection system. This may incubate the first practical side-channel-free QKD.

\section*{Methods}

\subsection*{GLLP security analysis with source flaws}
We discuss the standard GLLP security analysis for BB84 with source flaws~\cite{GLLP:2004,tamaki2012phase}, which is used for our simulation of Fig.~\ref{Fig:simulation:flaws}. We focus on phase encoding BB84 and assume \{$\delta_1$, $\delta_2$, $\delta_3$\} to be Alice's phase modulation errors for \{$\pi/2$, $\pi$, $3\pi/2$\}, thus the four BB84 imperfect states sent by Alice are given by
\begin{equation} \label{Eqn:state}
\begin{aligned}
\ket{\phi_{0_{z}}}&=\ket{0_{z}} \\
\ket{\phi_{1_{z}}}&=\sin\delta_2\ket{0_{z}}+\cos\delta_2\ket{1_{z}}\\
\ket{\phi_{0_{x}}}&=\cos\delta_1\ket{0_{x}}+\sin\delta_1\ket{1_{x}}\\
\ket{\phi_{1_{x}}}&=\sin\delta_3\ket{0_{x}}+\cos\delta_3\ket{1_{x}}
\end{aligned}
\end{equation}

Based on GLLP for imperfect sources, the $\esec$-secret key length is similar to Eqn.~\ref{eqn1}, except for the phase error rate, which includes the correction due to basis-dependent flaws and is revised to~\cite{GLLP:2004}
\be \label{eqn:phaseerror}
\bar{e}_{x,1}^{U}\leq e_{x,1}^{U}+4\Delta'+4\sqrt{\Delta'e_{x,1}^{U}}+\epsilon_{ph}
\ee
Here, $\Delta'$, called the balance of a quantum coin~\cite{GLLP:2004,tamaki2012phase}, quantifies the basis-dependent flaws of Alice signals associated with single-photon events. $\Delta'$ is given by~\cite{GLLP:2004}
\begin{equation} \label{Eqn:Deltap}
\begin{aligned}
\Delta' &\leq \frac{\Delta}{Y_1}\\
\Delta & = \frac{1-F(\rho_z,\rho_x)}{2}
\end{aligned}
\end{equation}
where $Y_1$ (typically called the yield of single photons~\cite{Lo:2005}) is the frequency of successful detections associated with single-photons; $F(\rho_z,\rho_x)$ is the fidelity of the density matrices for the $\mathsf{Z}$ and $\mathsf{X}$ basis. Using Eq.~(\ref{Eqn:state}), we can easily calculate $F(\rho_z,\rho_x)$ given \{$\delta_1$, $\delta_2$, $\delta_3$\}. In our QKD system, with \{$\delta_1$, $\delta_2$, $\delta_3$\} upper bounded by 0.127, we have $F(\rho_z,\rho_x)$=$1-1.9\times10^{-3}$. So, from Eq.~(\ref{Eqn:Deltap}), $\Delta$=$9.45\times10^{-4}$.

In GLLP analysis, the imperfect fidelity $F(\rho_z,\rho_x)$ can in principle be enhanced by Eve via exploiting the channel loss, which is clearly shown in Eq.~(\ref{Eqn:Deltap}), \ie, $\Delta$ is enhanced to $\Delta'$. Combined with the decoy-state estimations discussed in~\cite{lim2013concise}, we can derive the key length and obtain the inset curves in Fig.~\ref{Fig:simulation:flaws}.

\subsection*{Loss-tolerant protocol in a practical setting}
Our decoy-state analysis builds on~\cite{lim2013concise}, which discusses the decoy-state BB84. Our new contribution is estimating the phase error rate $e_{x,1}^{U}$. In decoy-state BB84, $e_{x,1}^{U}$ is estimated from the counts in $\mathsf{X}$ basis~\cite{lim2013concise}. In the loss-tolerant protocol, however, $e_{x,1}^{U}$ is estimated from the rejected counts, \ie, considering the detection events associated with single photons when Alice and Bob use different bases. Notice also that our estimation focuses directly on the detection \emph{counts} announced by Bob, which is different from previous analysis that is based on detection probabilities~\cite{Lo:2005, Wang:2005}.

In original decoy-state method~\cite{Lo:2005, Wang:2005}, Alice first randomly chooses an intensity setting (signal state or decoy state) to modulate each laser pulse and then she announces her intensity choices after Bob's detections. One can imagine a \emph{virtual} but equivalent protocol: Alice has the ability to first send $n$-photon states and then she only decides on the choice of intensity after Bob has a detection. Let $s_{z,n}$~be the number of detection counts observed by Bob given that Alice sends $n$-photon states in $\mathsf{Z}$ basis. Note that $\sum_{n=0}^\infty s_{z,n}=n_{z}$ is the total number of detections (gain counts). In the asymptotic limit with two decoy states, we have
\be \label{A:eqn:1} \nonumber
\hat{n}_{z,k}= \sum_{n=0}^\infty P_{k|n}s_{z,n},~\forall k \in \{\mu,\nu,\omega\},
\ee
where $P_{k|n}$ is the conditional probability of choosing the intensity $k$ given that Alice prepares an $n$-photon state. For finite-data size, from Hoeffding's inequality~\cite{Hoeffding}, the experimental measurement $n_{z,k}$ satisfies
\be \nonumber
\left|\hat{n}_{z,k}-n_{z,k} \right| \leq \Delta(n_z,\eps_1),
\ee
with probability at least $1-2\eps_1$, where $\Delta(n_z,\eps_1)=\sqrt{n_z/2\log(1/\eps_1)}$ and $\hat{n}_{z,k}$ is the expected value of $n_{z,k}$. Note that our analysis considers the most \emph{general} type of attack -- joint attack -- consistent with quantum memories. The above equation allows us to establish a relation between the asymptotic values and the observed statistics. Specifically,
\begin{eqnarray} \nonumber
 \hat{n}_{z,k} &\leq& n_{z,k}+\Delta(n_z ,\eps_1)=n_{z,k}^{U}, \\ \nonumber
 \hat{n}_{z,k} &\geq& n_{z,k}-\Delta(n_z ,\eps_1)=n_{z,k}^{L} ,
\end{eqnarray}
are respectively the upper and lower bound of the gain counts $n_{z,k}$ for a given intensity setting $k\in \{\mu,\nu,\omega\}$.

An analytical lower-bound on $s_{z,0}$ can be established by exploiting the structure of the conditional probabilities $P_{k|n}$ based on Bayes' rule: $P_{k|n}=\frac{P_{k}}{\tau_n}\frac{e^{-k}k^n}{n!}$, where $\tau_n=\sum_{k\in\{\mu,\nu,\omega\}}P_{k}e^{-k}k^n/n!$ is the probability that Alice prepares an $n$-photon state. Based on an estimation method in~\cite{ma2005practical}, we have
\be \label{A:eqn:4}
s_{z,0}^{L}=\frac{\tau_0}{(\nu-\omega)}\left(\frac{\nu e^{\omega} n^{L}_{z,\omega}}{P_{\omega}}-\frac{\omega e^{\nu}  n^{U}_{z,\nu}}{P_{\nu}}\right),
\ee
\begin{eqnarray} \label{A:eqn:5}
s_{z,1}^{L}= \frac{\mu \tau_1 }{\mu(\nu-\omega)-(\nu^2-\omega^2)}
\Bigg[  \frac{ e^{\nu}  n^{U}_{z,\nu}}{P_{\nu}}-\frac{ e^{\omega} n^{L}_{z,\omega}}{P_{\omega}} \\ \nonumber
+ \frac{\nu^2-\omega^2}{\mu^2} \left( \frac{s_{z,0}^{L}}{\tau_0}-\frac{e^{\mu}n^{U}_{z,\mu}}{P_{\mu}} \right) \Bigg].
\end{eqnarray}

Different from~\cite{lim2013concise}, the phase error rate $e_{x,1}^{U}$ in the loss-tolerant protocol is estimated using the rejected data analysis~\cite{tamaki2013loss}. That is, $e_{x,1}^{U}$ can be obtained by estimating the transmission rate of a fictitious (virtual) quantum signal sent by Alice. This transmission rate can be exploited by using the basis mismatch events. Supplementary Material shows the details for such an estimation in the asymptotic case. After considering the finite-data analysis, $e_{x,1}^{U}$ is given by

\begin{equation} \label{A:eqn4}
e_{x,1}^{U} =\frac{s_{0_x|1_x, 1}^{vir, U}+s_{1_x|0_x, 1}^{vir, U}}{s_{0_x|0_x, 1}^{vir, L}+s_{0_x|1_x, 1}^{vir, L} +s_{1_x|0_x, 1}^{vir, L}+s_{1_x|1_x, 1}^{vir, L}}.
\end{equation}
where
\begin{equation} \nonumber
\begin{bmatrix}P_{z}s_{0_x|j_x, 1}^{vir, \lambda}\\ P_{z}s_{1_x|j_x, 1}^{vir, \lambda}\end{bmatrix}=B\times A^{-1} \begin{bmatrix}2P_{x}s_{j_x|0_z, 1}^{\lambda} \\2P_{x}s_{j_x|1_z, 1}^{\lambda}\\ P_{z}s_{j_x|0_x, 1}^{\lambda}\end{bmatrix},
\end{equation}
where $P_{z}$ and $P_{x}$ are the probabilities that Alice and Bob choose $\mathsf{Z}$ and $\mathsf{X}$ basis, $j\in\{0, 1\}$, $\lambda\in\{U, L\}$ and A and B are given by (see Supplementary Material)
\begin{equation} \nonumber
\begin{split}
A&=\begin{bmatrix} 1& 1& 0 \\ 1& -\cos(2\delta_2)& \sin(2\delta_2)\\ 1& \sin(2\delta_1)& \cos(2\delta_1)\end{bmatrix}\\
B&=\frac{1}{12}\begin{bmatrix} (1+\sin\delta_2) &\sin\delta_2(1+\sin\delta_2) &\cos\delta_2(1+\sin\delta_2)\\ (1-\sin\delta_2)& - \sin\delta_2(1-\sin\delta_2)& - \cos\delta_2(1-\sin\delta_2)\end{bmatrix}.
\end{split}
\end{equation}
Here $s_{j_{x}|i_{z},1}^{U}$ ($s_{j_{x}|i_{z},1}^{L}$) denotes the upper (lower) bound of single-photon events when Bob has detections associated with bit ``j" in $\mathsf{X}$ basis, given that Alice sends a state of $i_{z}$ with $i\in\{0, 1\}$. $s_{j_{x}|i_{z},1}^{L}$ can be estimated equivalently by plugging $n_{j_{x}|i_{z},k}^{L}$ ($n_{j_{x}|i_{z},k}^{U}$) into Eqs.~(\ref{A:eqn:4}) and (\ref{A:eqn:5}) to replace $n_{z,k}^L$ ($n_{z,k}^U$). $s_{j_{x}|i_{z},1}^{U}$ can be estimated by
\begin{eqnarray} \nonumber
s_{j_{x}|i_{z},1}^{U} = \tau_{1}\frac{n_{j_{x}|i_{z},\nu}^{U}-n_{j_{x}|i_{z},\omega}^{L}}{\nu-\omega}.
\end{eqnarray}

\section*{Acknowledgments}
We thank M.~Curty, M. Jiang, C. Lim, M. Legr\'{e}, G. Ribordy, and particularly M. Lucamarini, for valuable discussions. Support from NSERC, the CRC program, Connaught Innovation fund, Industry Canada, the OGS Visa award, Air Force Office of Scientific Research (AFOSR), CryptoWorks21, US Office of Naval Research, Mike \& Ophelia Lazaridis Fellowship, the National Natural Science Foundation of China (Grant No. 61178010 and No. 11304391) and the China Scholarship Council (No. 201406470051) is gratefully acknowledged.

\newpage

\section*{Supplementary Material}

\subsection{Experimental counts}
In Table~\ref{Tab:exp:counts}, we list the raw experimental counts for each distance. In the 5 and 20 km experiments, we collected about 75 sets of data, with each set of about 104.5 million pulses sent out by Alice. This corresponds to a total number of pulses $N=7.84\times10^{9}$. In the 50 km experiment, we collected about 500 sets of data and sent a total number of $N=5.23\times10^{10}$ pulses. The experimental gain counts ($n_{z,k}$, $n_{x,k}$), error counts ($n_{e_z,k}$, $n_{e_x,k}$) and rejected counts ($n_{0_{x}|z,k}$, $n_{1_{x}|z,k}$) are listed in the Table.

\begin{table*}[ht!]
\centering
\begin{tabular}{c c  c  c  c  c  c}
  \hline \hline
   Distance & $n_{z,\mu}$ & $n_{z,\nu}$ & $n_{z,\omega}$ & $n_{x,\mu}$ & $n_{x,\nu}$ & $n_{x,\omega}$  \\
   \hline
   5km & $7.84\times10^{7}$ & $2.23\times10^{6}$ & $2.60\times10^{4}$ & $7.17\times10^{6}$ & $4.08\times10^{5}$ & $4.70\times10^{3}$  \\

   20km & $8.09\times10^{6}$ & $1.50\times10^{6}$ & $2.71\times10^{4}$ & $3.40\times10^{6}$ & $6.31\times10^{5}$ & $1.36\times10^{4}$ \\

   50km & $2.01\times10^{7}$ & $6.94\times10^{5}$ & $4.81\times10^{4}$ & $2.06\times10^{6}$ & $7.10\times10^{5}$ & $4.82\times10^{4}$ \\
  \hline

  & $n_{e_{z,\mu}}$ & $n_{e_{z,\nu}}$ & $n_{e_{z,\omega}}$ & $n_{e_{x,\mu}}$ & $n_{e_{x,\nu}}$ & $n_{e_{x,\omega}}$  \\
   \hline
   5km & $1.01\times10^{6}$ & $6.40\times10^{4}$ & $6.80\times10^{3}$ & $1.32\times10^{5}$ & $1.25\times10^{4}$ & $1.76\times10^{3}$  \\

   20km & $2.22\times10^{5}$ & $6.13\times10^{4}$ & $6.78\times10^{3}$ & $5.67\times10^{4}$ & $2.68\times10^{4}$ & $2.65\times10^{3}$ \\

   50km & $5.98\times10^{5}$ & $8.46\times10^{4}$ & $2.28\times10^{4}$ & $6.40\times10^{5}$ & $8.89\times10^{4}$ & $2.23\times10^{4}$ \\
   \hline

   & $n_{0_{x}|z,\mu}$ & $n_{0_{x}|z,\nu}$ & $n_{0_{x}|z,\omega}$ & $n_{1_{x}|z,\mu}$ & $n_{1_{x}|z,\nu}$ & $n_{1_{x}|z,\omega}$ \\
   \hline
     5km & $7.65\times10^{6}$ & $4.94\times10^{5}$ & $6.60\times10^{3}$ & $8.32\times10^{6}$ & $4.64\times10^{5}$ & $4.80\times10^{3}$  \\

   20km & $2.71\times10^{6}$ & $4.79\times10^{5}$ & $9.13\times10^{3}$ & $2.68\times10^{6}$ & $5.19\times10^{5}$ & $9.17\times10^{3}$ \\

   50km & $1.14\times10^{7}$ & $3.25\times10^{5}$ & $2.21\times10^{4}$ & $1.12\times10^{7}$ & $3.53\times10^{5}$ & $2.51\times10^{4}$ \\
  \hline \hline
\end{tabular}
\caption{\textbf{Experimental raw counts.} } \label{Tab:exp:counts}
\end{table*}

\subsection{Phase error rate in the asymptotic setting}
Here, we present the details of our method for phase error rate estimation. For simplicity, our discussion focuses on the asymptotic case, while the extended result for the finite-data case has been presented in the main text. The key idea is to apply the loss-tolerant protocol~\cite{tamaki2013loss} into our qubit model to estimate the phase error rate. Our qubit model, as defined in the main text, is given by Eq.~(\ref{Eqn:state}).

The phase error rate can be obtained by estimating the transmission rate of a \emph{fictitious} quantum signal sent by Alice~\cite{tamaki2013loss}. This transmission rate can be exploited by using the basis mismatch events (i.e., rejected-data
analysis).

\subsubsection{Definition}
The density matrices for the three encoding states $\ket{\phi_{0_{z}}}$, $\ket{\phi_{1_{z}}}$, $\ket{\phi_{0_{x}}}$ are:
\begin{equation}
\rho_{0_z}=|\phi_{0_{z}}\rangle\langle \phi_{0_{z}}|=(I+\sigma_z)/2,
\end{equation}

\begin{equation}
\begin{aligned}
\rho_{1_z} &=|\phi_{1_z}\rangle\langle \phi_{1_z}|=\begin{bmatrix}\sin^2\delta_2 & \sin\delta_2\cos\delta_2 \\
\sin\delta_2\cos\delta_2 & \cos^2\delta_2 \end{bmatrix} \\
& = \frac{1}{2}I - \frac{1}{2}\cos(2\delta_2) \sigma_z+\frac{1}{2}\sin(2\delta_2)\sigma_x,
\end{aligned}
\end{equation}

\begin{equation}
\begin{aligned}
\rho_{0_x} &=|\phi_{0_x}\rangle\langle \phi_{0_x}|=\frac{1}{2}\begin{bmatrix} 1+\sin(2\delta_1) & \cos(2\delta_1) \\ \cos(2\delta_1) & 1-\sin(2\delta_1)\end{bmatrix} \\
& = \frac{1}{2}I + \frac{1}{2}\sin(2\delta_1) \sigma_z+\frac{1}{2}\cos(2\delta_1)\sigma_x,
\end{aligned}
\end{equation}
Here $\sigma_{x,y,z}$ denote Pauli matrices and $I$ is identity matrix. The equivalent entanglement states between Alice and Bob are~\cite{tamaki2013loss}
\begin{equation}
\begin{aligned}
|\Psi_z\rangle& =(|0_z\rangle|\phi_{0_z}\rangle+|1_z\rangle|\phi_{1_z}\rangle)/\sqrt{2}\\
|\Psi_x\rangle& =|0_x\rangle|\phi_{0_x}\rangle.
\end{aligned}
\end{equation}

\subsubsection{Phase error rate estimation}

Let $Y_{s_\beta, j_\alpha}^{\omega}$ with $\omega\in\{Z, X\}$ and $s, j \in\{0, 1\}$  denote the joint probability that Alice (Bob) obtains a bit value j (s) conditional on the state preparation of $|\Psi_{\omega}\rangle$ and her (his) basis choice $\alpha$ ($\beta$), then the joint probabilities for different states are~\cite{tamaki2013loss}:

\begin{equation} \label{Eq:Ys0}
\begin{aligned}
Y_{s_x,0_z}^z & =\frac{2}{6}Tr[D_{sx} \sigma_{B,0_z}^z]= \frac{1}{6}Tr[D_{sx} \rho_{0_z}] \\
& =(q_{s_x|I}+q_{s_x|z})/6,
\end{aligned}
\end{equation}
where $\sigma_{B,0_z}^z=Tr_A[|0_z\rangle\langle0_z|\otimes I |\Psi_z\rangle\langle\Psi_z|]=\frac{1}{2}|\phi_{0_z}\rangle\langle\phi_{0_z}|$, and $q_{s_x|(I,x,z)}=Tr[D_{s_x}\sigma_{I,x,z}]/2$;

\begin{equation} \label{Eq:Ys1}
\begin{aligned}
Y_{s_x,1_z}^z & =\frac{2}{6}Tr[D_{sx} \sigma_{B,1_z}^z]= \frac{1}{6}Tr[D_{sx} \rho_{1_z}]\\
& =[q_{s_x|I}-\cos(2\delta_2)q_{s_x|z}+\sin(2\delta_2)q_{s_x|x}]/6,
\end{aligned}
\end{equation}
where $\sigma_{B,1_z}^z=Tr_A[|1_z\rangle\langle1_z|\otimes I |\Psi_z\rangle\langle\Psi_z|]=\frac{1}{2}|\phi_{1_z}\rangle\langle\phi_{1_z}|$;

\begin{equation} \label{Eq:Ys0x}
\begin{aligned}
Y_{s_x,0_x}^x & =\frac{1}{6}Tr[D_{sx} \sigma_{B,0_x}^x]= \frac{1}{6}Tr[D_{sx} \rho_{0_x}]\\
& =[q_{s_x|I}+\sin(2\delta_1)q_{s_x|z}+\cos(2\delta_1)q_{s_x|x}]/6,
\end{aligned}
\end{equation}
where $\sigma_{B,0_x}^x=Tr_A[|0_x\rangle\langle0_x|\otimes I |\Psi_x\rangle\langle\Psi_x|]=\frac{1}{2}|\phi_{0_x}\rangle\langle\phi_{0_x}|$.

Eqs.~(\ref{Eq:Ys0})-(\ref{Eq:Ys0x}) can be rewritten as
\begin{equation} \label{Eq:YYY}
\begin{aligned}
\begin{bmatrix}Y_{s_x,0_z}^z \\Y_{s_x,1_z}^z\\ Y_{s_x,0_x}^x\end{bmatrix}
& = \frac{1}{6}\begin{bmatrix}Y_{s_x|0_z}^z \\Y_{s_x|1_z}^z\\ Y_{s_x|0_x}^x\end{bmatrix} = \frac{1}{6}\begin{bmatrix} 1& 1& 0 \\ 1& -\cos(2\delta_2)& \sin(2\delta_2)\\ 1& \sin(2\delta_1)& \cos(2\delta_1)\end{bmatrix} \begin{bmatrix} q_{s_x|I}\\ q_{s_x|z}\\ q_{s_x|x}\end{bmatrix} \\
& \equiv \frac{1}{6} A \begin{bmatrix} q_{s_x|I}\\ q_{s_x|z}\\ q_{s_x|x}\end{bmatrix}.
\end{aligned}
\end{equation}
Here $Y_{s_x|0_z}^z$ denotes the conditional probability that Bob obtains bit $s$ in basis $x$ given that Alice sends $0_z$. The same definition is applied to $Y_{s_x|1_z}^z$ and $Y_{s_x|0_x}^x$. Note that all these quantities can be measured \emph{directly} in experiment.

To estimate the phase error rate, we consider a \emph{virtual} protocol: Alice first prepares
$|\Psi_z\rangle$ and then both Alice and Bob measure systems A and B in the $X$ basis~\cite{tamaki2013loss}. The joint probabilities of the virtual states $Y_{s_x,j_x}^{z,vir}$ are:

\begin{widetext}
\begin{equation} \label{Eq:YY}
\begin{aligned}
Y_{s_x,0_x}^{z,vir}& =\frac{1}{12}Tr[D_{s_x}\sigma_{B,0_x}^{z,vir}]= \frac{1}{3} [(1+\sin\delta_2)q_{s_x|I}+ \sin\delta_2(1+\sin\delta_2)q_{s_x|x}+ \cos\delta_2(1+\sin\delta_2) q_{s_x|x}],\\
Y_{s_x,1_x}^{z,vir}& =\frac{1}{12}Tr[D_{s_x}\sigma_{B,1_x}^{z,vir}]= \frac{1}{3} [(1-\sin\delta_2)q_{s_x|I}- \sin\delta_2(1-\sin\delta_2)q_{s_x|x}- \cos\delta_2(1-\sin\delta_2) q_{s_x|x}].\\
\end{aligned}
\end{equation}
\end{widetext}

Eq.~(\ref{Eq:YY}) can be rewritten as
\begin{widetext}
\begin{equation}
\begin{bmatrix}Y_{s_x,0_x}^{z,vir}\\ Y_{s_x,1_x}^{z,vir}\end{bmatrix}= \frac{1}{12}\begin{bmatrix} (1+\sin\delta_2) &\sin\delta_2(1+\sin\delta_2) &\cos\delta_2(1+\sin\delta_2)\\ (1-\sin\delta_2)& - \sin\delta_2(1-\sin\delta_2)& - \cos\delta_2(1-\sin\delta_2)\end{bmatrix} \begin{bmatrix} q_{s_x|I}\\ q_{s_x|z}\\ q_{s_x|x}\end{bmatrix}\equiv B \begin{bmatrix} q_{s_x|I}\\ q_{s_x|z}\\ q_{s_x|x}\end{bmatrix}.
\end{equation}
\end{widetext}

Combining it with Eq.~(\ref{Eq:YYY}), we can obtain the rate of virtual states based on experimental results, which is
\begin{equation}
\begin{bmatrix}Y_{s_x,0_x}^{z,vir}\\ Y_{s_x,1_x}^{z,vir}\end{bmatrix}=B\times A^{-1} \begin{bmatrix}Y_{s_x|0_z}^z \\Y_{s_x|1_z}^z\\ Y_{s_x|0_x}^x\end{bmatrix}.
\end{equation}

Finally the phase error can be estimated by
\begin{equation} \label{Eq:ex}
e_x=\frac{Y_{1_x,0_x}^{z,vir}+Y_{0_x,1_x}^{z,vir}}{Y_{0_x,0_x}^{z,vir}+Y_{1_x,0_x}^{z,vir} +Y_{0_x,1_x}^{z,vir}+Y_{1_x,1_x}^{z,vir}}.
\end{equation}
The extended result of Eq.~(\ref{Eq:ex}) for the finite-data case is presented in the Methods of the main text.

\subsection{Qubit assumption}
We verify the qubit assumption, \ie, that the four BB84 states remain in two dimensions. This assumption is commonly made in various QKD protocols including decoy-state BB84 and MDI-QKD. We focus on a standard \emph{one-way phase-encoding} system, which has been widely implemented in experiments~\cite{rosenberg2007long,dixon2008gigahertz,lucamarini2013efficient,Patel2014}. In this system, LiNbO$_{3}$ waveguide-based phase modulator (PM) is commonly used to encode/decode phase information. Fig.~\ref{Fig:PM} illustrates the schematic of such PM~\cite{Yariv:2007}.  For commercial products, see~\cite{PM:Commercial}. To guarantee the qubit assumption, Alice's PM is supposed to have the same timing, spectral, spatial and polarization mode information for different BB84 states. We find that timing and spatial information can be easily guaranteed without any additional devices, while spectral and polarization information can also be guaranteed with standard low-cost optical devices such as wavelength filter and polarizer. Therefore, based on standard devices, we can verify the qubit assumption with high accuracy. We remark that our method serves as a specific example to practically verify the qubit assumption. In future, it will be interesting to work towards constructing a more general theory on the verification of the qubit assumption.

In the following, we discuss timing, spectral, spatial and polarization properties for different encoding phases.

\subsubsection*{Temporal-spectral mode}

\begin{figure}[!t]
\centering
\resizebox{8.5cm}{!}{\includegraphics{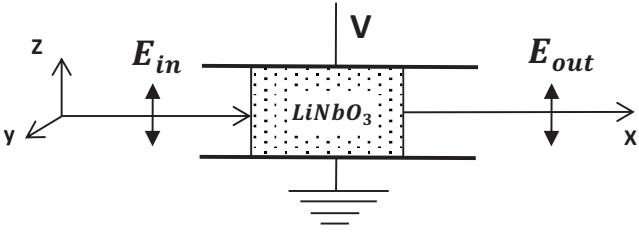}} \caption{Schematic of an electro-optic phase modulator based on LiNbO$_{3}$ crystal~\cite{Yariv:2007}. Commercial products can be seen in~\cite{PM:Commercial}. The double-headed arrows show the direction of polarization of the optical beam. The crystal is cut in a configuration so that the applied electrical field (voltage) is along the direction of the principal ($z$) axis. To take the advantage of the largest electro-optical coefficient in the $z$ axis, an optical beam is propagating along the $x$ axis, with the direction of polarization parallel to the $z$ axis.} \label{Fig:PM}
\end{figure}

\emph{Temporal mode:} Fig.~\ref{Fig:PM} shows the schematic of the phase modulation based on LiNbO$_{3}$ crystal. When phase modulator (PM) modulates different phases, the electrical-optical effect inside the LiNbO$_{3}$ waveguide changes the principal refractive index $n_z$. At first sight, it might appear that the timing information is indeed changed for different phase modulations. However, we will show that such change is so small that it can be neglected.

According to the EM theory in LiNbO$_{3}$ waveguide, the relations among the principal refractive index $n_z$, the group refractive index $n_g$ and the extraordinary refractive index $n_e$ are given by~\cite{Yariv:2007}
\begin{equation} \label{Eqn:Index}
\begin{aligned}
n_g & = n_z+\omega_0\frac{dn_z(\omega)}{d\omega}|_{\omega_0}\\
n_z & = n_e-\frac{1}{2}n_e^3r_z\frac{V}{d}
\end{aligned}
\end{equation}
where $\omega_0$ is the central frequency of the optical field, $r_z$ is the electro-optical coefficient along $z$ axis, $V$ is the voltage applied onto the crystal, and $d$ is the thickness of the crystal. Thus the timing difference $\Delta t$ between \{0\} and phase modulation \{$\pi$\} is given by
\begin{equation} \label{Eqn:deltat}
\begin{aligned}
\Delta t=[\frac{1}{2}n_e^3r_z\frac{V_{\pi}}{d}+\frac{3}{2}n_e^2r_z\frac{V_{\pi}}{d}\omega_0\frac{dn_e(\omega)}{d\omega}|_{\omega_0}]\frac{l_0}{c}
\end{aligned}
\end{equation}
where $V_{\pi}=\frac{\lambda_{0}d}{n_e^2r_zl_0}$ is the half-wave voltage that provides a phase modulation \{$\pi$\}~\cite{Yariv:2007}, $l_{0}$ is the length of the crystal and $c$ is the speed of light.

For a typical LiNbO$_{3}$ crystal working in the telecom wavelength $\lambda_{0}\sim 1550$ nm, it is well known that the relation between $n_e$ and $\lambda_{0}$ is given by~\cite{zelmon1997infrared}
\begin{equation} \label{Eqn:ne}
\begin{aligned}
n_e^2=1+\frac{2.980\lambda_{0}^2}{\lambda_{0}^2-0.020}+\frac{0.598\lambda_{0}^2}{\lambda_{0}^2-0.067}+\frac{8.954\lambda_{0}^2}{\lambda_{0}^2-416.08}
\end{aligned}
\end{equation}
Notice that in a waveguide based PM, one has to use the effective index, \ie, $n_{eff}$, to include the waveguide effect. We remark however that, for LiNbO$_{3}$ material, $n_{eff}$ and $n_e$ are almost the same~\cite{suhara2003waveguide}. Hence, by plugging Eq.~(\ref{Eqn:ne}) into Eq.~(\ref{Eqn:deltat}), we have $\Delta t\approx 4\times10^{-6}$ ns. In a QKD implementation, the optical pulse is typically around 1 ns width~\cite{rosenberg2007long,peng2007experimental,schmitt2007experimental} or 0.1 ns~\cite{dixon2008gigahertz,lucamarini2013efficient,Patel2014}, thus $\Delta t\ll 0.1$ ns. Assuming that the optical pulse is Gaussian, $\Delta t$ corresponds to a fidelity of $F(\rho^{0},\rho^{\pi})\approx1-10^{-8}$ between \{0\} and \{$\pi$\}. Therefore, timing remains (almost) the same for different phase modulations.

\emph{Spectral mode:} First, in a standard one-way system, Alice can locally synchronize the devices so that the optical pulse passes through Alice's PM in the middle of the electrical modulation signal (flat response). Hence, the optical pulse experiences a correct modulation \emph{without} spectral change~\cite{Xu:phaseremapping:2010,jiang2014frequency}. In a two-way system, Alice can monitor the timing information between the signal and reference pulse to guarantee the correct modulation and defend against side-channel attacks~\cite{Xu:phaseremapping:2010,jiang2014frequency}. Second, to guarantee single spectral mode from the output of a laser, one can use a standard wavelength filter. For instance, a recent QKD experiment used an off-the-shelf wavelength filter with a full-width at the half maximum (FWHM) of $\Delta\nu=$15 GHz for a different purpose~\cite{Patel2014}. In this case, given a Gaussian pulse with FWHM $\Delta t=$ 0.1 ns in the time domain~\cite{Patel2014}, it is quite close to the lower bound of time-bandwidth product~\cite{Yariv:2007}, \ie, $\Delta t \times \Delta \nu \geq \frac{2ln2}{\pi}$. Wavelength filters with narrow bandwidth have already been widely available on the market~\cite{AWF}. Hence, single spectral mode can be guaranteed with high accuracy by using a wavelength filter.

\subsubsection*{Spatial mode}
For a standard single-mode fiber (SMF), the core diameter is around 10 um. Theory and experiments have already confirmed that a SMF in the telecom wavelength rejects all high-order modes and conducts only one fundamental transverse mode~\cite{mynbaev2001fiber}. The cutoff wavelength of a standard SMF is about 1260 nm~\footnote{See, for instance, Corning's SMF28: http://www.corning.com/docs/opticalfiber}. Using the software of BeamPROP, we have also performed a numerical simulation with a standard multi-mode fiber propagating into a SMF. The results show that after only about one millimeter, SMF rejects almost all high-order modes. The high-order modes decay exponentially, thus after about ten millimeters, there is no high-order component left (less than $10^{-10}$ proportion). Notice that, the input of a standard commercial PM usually has a certain length of pigtail fiber (about one meter)~\cite{PM:Commercial}. Therefore, the single mode assumption on spatial mode can be easily guaranteed in practice.

\subsubsection*{Polarization mode}
The input of a commercial PM is normally a pigtail of polarization maintaining fiber~\cite{PM:Commercial}, which can ensure that the input polarization is perfectly aligned with the principal axis of PM. Experimentally, before this polarization maintaining fiber, one can use a fiber polarization beam splitter (PBS) to reject other polarization modes. A standard PBS has about 30 dB extinction ratio. In the following, we discuss the error due to this finite extinction ratio (30 dB). Ideally, if the PBS has infinite extinction ratio, the input state is perfectly aligned with the principal axis ($z$ axis in Fig.~\ref{Fig:PM}) and Alice modulates the four BB84 states as
\begin{equation} \nonumber
\begin{aligned}
\ket{\phi_{j}}&=\frac{1}{\sqrt{2}}(e^{ij\frac{\pi}{2}}\ket{S_z}+\ket{R_z}),
\end{aligned}
\end{equation}
where $j\in$\{0, 1, 2, 3\} denotes the four BB84 states and $\ket{S_z}$ ($\ket{R_z}$) denotes the signal (reference) pulse with polarization along $z$ axis. However, due to the finite extinction ratio of PBS, the signal and reference pulse are expressed as
\begin{equation} \nonumber
\begin{aligned}
\ket{S}&=\alpha\ket{S_y}+\beta\ket{S_z},\\
\ket{R}&=\alpha\ket{R_y}+\beta\ket{R_z},
\end{aligned}
\end{equation}
where $\ket{S_y}$ denotes the polarization component along $y$ axis. For 30 dB extinction ratio, $\alpha^2\approx$0.001. Thus Alice's imperfect modulations can be described by
\begin{equation} \label{Eqn:imperfectBB84state}
\begin{aligned}
\ket{\phi'_{j}}&=\frac{1}{\sqrt{2}}(\alpha e^{ij\frac{\pi}{6}}\ket{S_y}+\beta e^{ij\frac{\pi}{2}}\ket{S_z}+\alpha\ket{R_y}+\beta\ket{R_z},
\end{aligned}
\end{equation}
where we assume that the relative modulation magnitude ratio between the polarization aligned with the principal axis ($z$ axis) and the orthogonal polarization ($y$ axis in Fig.~\ref{Fig:PM}) is 1:3~\cite{Yariv:2007,Xu:phaseremapping:2010}. Using three new bases \{$\ket{e_1}$, $\ket{e_2}$, $\ket{e_3}$\}, Eq.~(\ref{Eqn:imperfectBB84state}) can be written as (similar to~\cite{sun2011passive})
\begin{equation}\label{Eqn:imperfectBB84state2}
\begin{aligned}
\ket{\phi'_{j}}=\frac{1}{\sqrt{2}}(\alpha\beta(e^{ij\frac{\pi}{6}}-e^{ij\frac{\pi}{2}})\ket{e_1}+(\alpha^2e^{ij\frac{\pi}{6}}+\beta^2e^{ij\frac{\pi}{2}})\ket{e_2}
+\ket{e_3}),
\end{aligned}
\end{equation}
Hence, the four imperfect states is spanned to three dimensions in Hilbert space, \ie, the information encoded by Alice is not only in the time-phase mode but also in the polarization mode. However, for 30 dB extinction ratio, we find that it is almost impossible for Eve to attack the system, because the fidelity between $\ket{\phi_{j}}$ and $\ket{\phi'_{j}}$, $F(\rho^{\ket{\phi_{j}}},\rho^{\ket{\phi'_{j}}})=tr \sqrt{\sqrt{\rho^{\ket{\phi_{j}}}}\rho^{\ket{\phi'_{j}}}\sqrt{\rho^{\ket{\phi_{j}}}}}$, is about $1-10^{-7}$ for $j\in$\{0, 1, 2, 3\}. This shows that the imperfect states are highly close to the perfect BB84 states. Most importantly, one can derive a refined security proof to include this small imperfection into the secure key rate formula, which will be a subject of future investigation.

\bibliographystyle{apsrev}

\end{document}